\documentclass[%
reprint,
superscriptaddress,
sd,
amsmath,amssymb,
nofootinbib,
]{revtex4-1}
\usepackage{color}
\usepackage{hyperref}
\usepackage{float}
\usepackage{graphicx}% Include figure files
\usepackage{textcomp}
\usepackage{epsfig}
\usepackage{epstopdf}
\usepackage{natbib}
\usepackage{multirow}
\usepackage{siunitx}
\usepackage{braket}
\usepackage[titletoc]{appendix}
\usepackage{dcolumn}
\usepackage{bm}% bold math
\newcommand{\affilWuerz}{Technische Physik, Universit\"{a}t W\"{u}rzburg, Am Hubland, 97074 W\"{u}rzburg, Germany}
\newcommand{\affilInnsbruck}{Institut f\"{u}r Experimentalphysik, Universit\"{a}t Innsbruck, Technikerstra{\ss}e 25, 6020 Innsbruck, Austria}

\begin{document}
\preprint{APS/123-QED}

\title{Understanding photoluminescence in semiconductor Bragg-reflection waveguides: Towards an integrated, GHz-rate telecom photon pair source}% Force line breaks with \\
\author{S. Auchter}
\affiliation{\affilInnsbruck}%
\affiliation{Infineon Technologies Austria AG, Siemensstra{\ss}e 2, 9500 Villach, Austria}
\author{A. Schlager}
\email[]{alexander.schlager@uibk.ac.at}
\affiliation{\affilInnsbruck}%
\author{H. Thiel}
\affiliation{\affilInnsbruck}%
\author{K. Laiho}
\affiliation{Institute of Quantum Technologies, German Aerospace Center (DLR), S\"{o}flinger Straße 100, 89077 Ulm, Germany}%
\author{B. Pressl}
\affiliation{\affilInnsbruck}%
% \affiliation{Besi Austria GmbH, Innstraße 16, 6241 Radfeld, Austria}
\author{H. Suchomel}
\affiliation{\affilWuerz}%
\author{M.~Kamp}
\affiliation{\affilWuerz}%
\author{S.~H\"{o}fling}
\affiliation{\affilWuerz}%
\affiliation{School of Physics \& Astronomy, University of St Andrews, St Andrews KY16 9SS,~UK}%
\author{C.~Schneider}
\affiliation{\affilWuerz}%
\affiliation{Institute of Physics, University of Oldenburg, D-26129 Oldenburg, Germany}
\author{G.~Weihs}
\affiliation{\affilInnsbruck}%

%\affiliation{Department of Physics, National University of Defense Technology, Changsha, 410073 People’s Republic of China}
%\affiliation{Institut f\"{u}r Experimentalphysik, Universit\"{a}t Innsbruck, Technikerstra{\ss}e 25, 6020 Innsbruck, Austria}%
%
%    \author{Gregor Weihs}%
%    \email{gregor.weihs@uibk.ac.at}    %
%    \affiliation{Institut f\"{u}r Experimentalphysik, Universit\"{a}t Innsbruck, Technikerstra{\ss}e 25, 6020 Innsbruck, Austria}

%\author{AA}
% \author{BB}%
%  \author{CC}%
%   \author{DD}%
   %\email{benedikt.pressl@uibk.ac.at, benedikt.pressl@gmail.com}
%

\begin{abstract}
Compared to traditional nonlinear optical crystals, like BaB$_2$O$_4$, KTiOPO$_4$ or LiNbO$_3$, semiconductor integrated sources of photon pairs may operate at pump wavelengths much closer to the bandgap of the materials. This is also the case for Bragg-reflection waveguides (BRW) targeting parametric down-conversion (PDC) to the telecom C-band. The large nonlinear coefficient of the AlGaAs alloy and the strong confinement of the light enable extremely bright integrated photon pair sources. However, under certain circumstances, a significant amount of detrimental broadband photoluminescence has been observed in BRWs. We show that this is mainly a result of linear absorption near the core and subsequent radiative recombination of electron-hole pairs at deep impurity levels in the semiconductor. For PDC with BRWs, we conclude that devices operating near the long wavelength end of the S-band or the short C-band require temporal filtering shorter than \SI{1}{\nano\second}. We predict that shifting the operating wavelengths to the L-band and making small adjustments in the material composition will reduce the amount of photoluminescence to negligible values. Such measures enable us to increase the average pump power and/or the repetition rate, which makes integrated photon pair sources with on-chip multi-gigahertz pair rates feasible.
\end{abstract}

% PACS, the Physics and Astronomy  Classification Scheme.
%42.65.Wi 	Nonlinear waveguides
%42.65.Lm 	Parametric down conversion and production of entangled photons
%78.55.Cr   Photoluminescence, properties and materials: III-V semiconductor

%Use showkeys class option if keyword
\date{\today}%

\maketitle

%----------------------------------------------------------------------------------
\section{Introduction}
%----------------------------------------------------------------------------------

%OLD: Robust, entangled photon sources are vital for performing quantum optics tasks fast and reliably. For example in a variety of quantum communication applications, no matter whether performed on the ground or via a satellite  \cite{Rudolph-Why-optimistic-2016, Ma2012,Yin-Pan-Satellite-2017,Guenthner2017}, integrated quantum resources can turn very useful. If the entangled photon sources that are often based on bulk crystals are replaced with integrated optics, one not only greatly saves space and reaches better scalability but also gains in optical stability \cite{brien--science--2007,politi--science--2008}.

Entangled photons form the basis of many quantum applications, notably in computing and communication \cite{brien--science--2007, Rudolph-Why-optimistic-2016, Ma2012, Guenthner2017}. For this purpose, one would like to have sources that produce single photons, photon pairs or entangled photons at high rates with high quality. In practical realizations, various technological challenges arise which affect the efficiency or quality of the prepared photon states. This can range from the simple dark noise in a detector to complex, parasitic nonlinear optical interactions in an optical component. 

While some strategies to handle and mitigate detrimental effects are often internal lab-knowledge (e.g. the ubiquitous black masking tape at the right places to shield detectors from background light), others are well-known techniques and approaches in the experiment or data post-processing. For example, in both bulk and integrated experiments, pulsed operation, time gating, and spectral and spatial filtering are commonly employed \cite{aspect1981,kwiat1995,patel2012,eraerds2010,schweickert2018}. 

To be more specific, many quantum optics experiments are plagued by uncorrelated background light that produces spurious events and reduces the quality of the photon state and the signal-to-noise ratio \cite{shields2007}. This is not necessarily limited to the photon pair sources: In quantum cryptography applications, for example, light leakage can lead to compromised link security \cite{brassard2000}. 

Integrated semiconductor quantum light sources are particularly susceptible to the parasitic influence of background light. Once the photonic chips have been fabricated, there are only limited options to include additional filters afterwards. Moreover, the presence of imperfections in semiconductor materials causes many complex light-matter interactions that are difficult to track down or get rid of.

Nevertheless, integrated photonic circuits have the advantage of overall stability and compact dimensions. Small dimensions lead to high field strengths which increase nonlinear interactions. The further integration of light sources drastically improves the wall-plug efficiency of these chips, compared to bulk setups. This is all beneficial for many potential practical applications \cite{brien--science--2007, politi--science--2008}, but especially for satellite technology \cite{armengol2008, Yin-Pan-Satellite-2017}. Thus, it is worthwhile to look at measures to suppress the parasitic effects already at the source level on the chip.

One example of integrated quantum light sources are Bragg-reflection waveguides (BRWs) that produce photon pairs via parametric down-conversion (PDC) \cite{lanco2006semiconductor,sarrafi2013continuous,Gregor-Monolithic-Source-2012}. They are made of multiple epitaxial layers of different alloys of aluminum gallium arsenide (AlGaAs). This material system possesses a large second-order optical nonlinearity \cite{Boyd2003} and is versatile for designing and fabricating samples with the help of well-established techniques. A great benefit of AlGaAs is its potential to seamlessly integrate electro-optic elements, like light sources and modulators, with PDC on chip. The waveguides can be designed to operate at almost any temperature where the material is stable. While the technology still is not as mature as silicon or lithium niobate in certain aspects, most notably linear loss and homogeneity, it is improving quickly \cite{Pressl--2015, porkolab2014algaasreflow, Pressl-2017-advanced-BRW}.

Recently, BRWs have become increasingly popular for PDC \cite{Gregor-Monolithic-Source-2012,Gunthner-2015,Claire2016Multi-user,kang2016monolithic}, and considerable effort was dedicated towards optimizing their performance for different tasks. For example, not only integrated electrically injected pump lasers on the same chip as the PDC sources could be demonstrated \cite{Boitier-Ducci-Electrically-2013,bijlani2013semiconductor}, but also various aspects of the preparation of polarization entangled states \cite{Valles-2013,Horn-scientific-reports-2013,kang2015two}, like the compensation of the birefringent group delay \cite{schlager--2017--temporally}.

In this paper, we conduct a series of experiments to determine the driving factors that affect the generation of photoluminescence in our BRWs, such as wavelength and power dependencies as well as the spatial field distribution. Photoluminescence in BRW structures similar to the ones investigated here has been reported by other groups \cite{Horn-scientific-reports-2013,Boitier-Ducci-Electrically-2013} as well, but was never studied in-depth. It was informally hypothesized that the impurities in the substrate are driving factors of the photoluminescence. 

We note that such luminescence is not exclusive to PDC in semiconductors, but also exists in commonly employed crystals, like BaB$_2$O$_4$ or KTiOPO$_4$, especially when pumped with high-energy light \cite{machulka-2014-bbo-luminescence,Bhar-89-evaluation-bbo,Chen-09-pdc-ppktp}. At similar wavelengths to our BRWs, fluorescence is observed in periodically poled silica fibers \cite{Zhu-11-pp-fiber} and spontaneous Raman scattering in four-wave-mixing schemes (FWM) \cite{Takesue-05-sfwm-fiber}. In integrated quantum optics, parasitic background light has also been observed in p-i-n diode based single-photon sources \cite{yuan2002} and quantum-dot based systems \cite{stevenson2006}.

\subsection{Aspects of photoluminescence in integrated PDC sources}

The challenge of parasitic photoluminescence in integrated PDC sources of photon pairs is that they operate in the linear low-gain regime \cite{christ2013}. In contrast to classical second-harmonic generation, difference frequency generation or optical parametric oscillation schemes, the efficiency \textit{does not} increase with increasing pump power. Typical PDC signal rates are in the kHz to GHz range which are many orders of magnitude smaller than the photon rate of the pump light ($\gg$\si{\peta\hertz}). These drastically different amounts of optical power in the device mean that even very inefficient photoluminescence processes can easily produce photons at rates similar to those of the PDC. 

The observed photoluminescence rates depend on the power nonlinearity of the underlying process: If the PDC is pumped with a pulsed laser incorporating a high peak power, nonlinear photoluminescence generation is either greatly enhanced or somewhat suppressed, compared to the continuous wave (CW) case. The former corresponds to unrestricted two- or multi-photon absorption, while the latter indicates saturation. 

PDC in BRWs is based on the interaction between the fundamental and higher-order spatial modes. One challenging aspect in this regard is that BRWs do not operate in a single-mode regime, but support higher-order modes with different mode profiles. These characteristic profiles are the result of the stratified layout made from different material compositions and the horizontal confinement of the ridge. Hence, the layers are exposed to different amounts of light depending on whether we look at the pump or PDC wavelength. For example, the pump mode is localized in the central core layer, while the fundamental telecom mode of the PDC photons mostly propagates in the two layers right adjacent to the core. Of course, the material composition of these is different from the core, which results in additional complications when analysing the inter-play of pump light, PDC and the photoluminescence.

Due to the complicated modal structure, it is easy to excite many spatial modes simultaneously, not always intentionally. Therefore, a considerable amount of light can excite the semiconductor material, but will not partake in down-conversion \cite{Pressl--2015}. There are also many modes which not only excite the impurities, but will guide any photoluminescence along the ridge. These conditions are sub-optimal as the PDC also propagates in the waveguide. Various external improvements have been proposed to reduce the effective multi-modeness, for example beam-shaping via holographic elements or integrating the pump laser \cite{Boitier-Ducci-Electrically-2013,bijlani2013semiconductor}. Narrow temporal filtering below \SI{3}{ns} is commonly employed, and we have previously shown that additional spectral filtering is highly effective \cite{Gunthner-2015}.

In this work, however, we \textit{intentionally ignore} these \textit{best practices} in order to get a clear photoluminescence signal. We use an excitation laser with a Gaussian beam profile and use only wide time-gates of \SI{13}{\nano\second} (= pulse repetition time of the laser). This allows us to model the photoluminescence from studying the single and coincidence rates with varying frequency and incident power. One goal is to separate the contribution of the photoluminescence from the PDC. In this context, we also study whether the specific design of the sample can be improved to reduce noise. 

The data presented here reveals that the photoluminescence results from linear absorption at the bandgap of one or more layers near the core of the waveguide. Once an electron-hole pair is excited, the radiative recombination takes place at impurities at half the bandgap energy. Photoluminescence from these deep levels can be related to lattice defects, like antisites (arsenic) or vacancies, or certain rare dopants \cite{pavesi-1994-photoluminescence}. Moreover, we operate our device at room-temperature, therefore a rather broad, quasi-uniform photoluminescence spectrum is expected \cite{pavesi-1994-photoluminescence}. This is consistent to measurements in previous experiments \cite{laiho--2016--uncovering}.

%----------------------------------------------------------------------------------
\section{Methods}
%----------------------------------------------------------------------------------
\subsection{Sample and setup}
\begin{figure}[ht]
\centering
\includegraphics[width=\columnwidth]{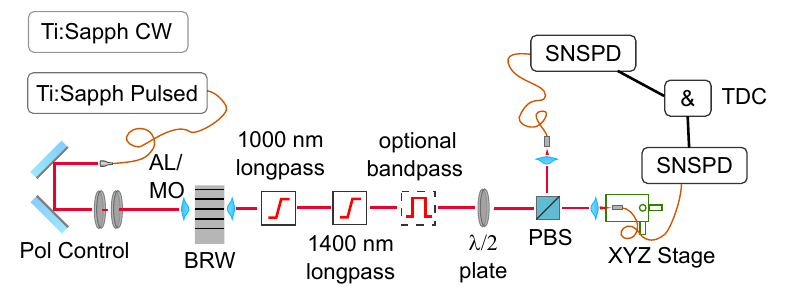}
\caption{\label{fig:main-setup} Schematic of the main setup for the pulsed excitation. The selected pump laser is coupled into the waveguide via an aspheric lens (AL) or a microscope objective (MO) on one side and the generated PDC is collected with another AL at the output facet. After collimation, the pump is suppressed by two longpass filters, followed by an optional bandpass. The photon pair is split deterministically at the polarizing beam splitter (PBS) and coupled to fibers via ALs on the XYZ translation stages and detected and correlated subsequently.}
\end{figure}

The BRW sample under investigation is a state-of-the-art, low-loss, matching-layer enhanced design optimized for simple fabrication, while simultaneously allowing a bright type\nobreakdash-II PDC process \cite{Pressl-2017-advanced-BRW}. Our experimental setup is shown in Fig.~\ref{fig:main-setup}. For illumination, we switch between two different pump lasers. The first is a Coherent MIRA titanium sapphire laser running in femtosecond mode, with the pulses being stretched to \SI{1.5}{ps} by a pulse stretcher. The pulse repetition rate is \SI{76.2}{MHz}, which corresponds to a \SI{13.1}{ns} delay between pulses. The second is a continuous wave Tekhnoscan T\&D\nobreakdash-scan X1 titanium sapphire laser. The selected pump laser is coupled into the waveguides via an aspheric lens (AL) or a microscope objective (MO) on one side, the generated PDC is collimated with another AL at the output facet. Two longpass dichroic interference filters remove the residual pump beam. An optional \SI{40}{nm} or \SI{12}{nm} bandpass filter, nominally centered at \SI{1550}{nm}, follows before the photon pairs are split by a polarizing beam splitter (PBS). The central wavelength of the bandpass is slightly tunable by rotating the filter a few degrees with respect to the beam. After the PBS, we couple the photons via collimation optics into single mode fibers connected to high-efficiency SingleQuantum EOS superconducting nanowire single photon detectors (SNSPDs). The output of the SNSPDs is amplified, routed to threshold discriminators and detected by a quTools quTau time\nobreakdash-to\nobreakdash-digital converter (TDC). All optics have the necessary broadband coating and suitable glass. This ensures a reflectivity of less than \SI{1}{\percent} at a \SI{100}{nm} away from the nominal operating wavelengths.

We conduct three experiments: The first in section~\ref{sec:spatial} serves as a sanity check, testing the spatial distribution of the PDC and photoluminescence signals at the facet. Here, we move the collection fiber (single\nobreakdash-mode) parallel to the facet to verify that both signals are indeed coming from the waveguide. The second experiment in section~\ref{sec:rate-model} yields a precise model of the PDC and noise. We vary the pump power and employ various bandpass filters after the waveguide while recording the channel rates and coincidences. The final experiment in section~\ref{sec:off-resonant} reveals more about the cause of the photoluminescence by measuring the signal power at pump wavelength strongly detuned from the degeneracy point.
 
For convenience, we operate two almost identical setups according to Fig.~\ref{fig:main-setup}. In sections~\ref{sec:spatial} and \ref{sec:off-resonant} we employ  a waveguide with a length of \SI{2.04}{\milli\meter} and a degeneracy wavelength of \SI{767}{nm} as well as a microscope objective for the in-coupling. In sections~\ref{sec:rate-model} and \ref{sec:future-devices} a \SI{1.3}{\milli\meter} long waveguide with \SI{763}{\nano\meter} degeneracy wavelength is utilized and an aspheric lens is employed for in-coupling. To ensure comparability, both are from the same wafer. These wavelengths are shorter than the design wavelength of \SI{775}{\nano\metre}. Depending on the location of the chip on the wafer and the waveguide width some variability was observed. We intentionally choose waveguides with shorter operating wavelengths, as they are closer to the bandgap of the material and yield a clearer combined PDC-photoluminescence signal as investigated in section~\ref{sec:off-resonant}.

The main experimental challenge is keeping all light sources and the setup stable under comparable and reproducible conditions. This imposes a practical limit how much usable data we can actually acquire in a certain time-frame. As some components, like filters, need to be changed for every set, the coupling varies and quick re-alignment is necessary. Therefore, we optimize the setup on metrics that can be derived in real-time in the lab\footnote{With a more time-consuming and elaborate optimization procedure, values similar to better than the ones reported in Ref.~\cite{Gunthner-2015} can readily be achieved (e.g. Ref.~\cite{chen-2018-time-bin})}, most importantly the raw coincidence count rate. We operate the setup in a tightly controlled environment suitable for precision interferometry as reported in our Ref.~\cite{Kauten2017}.

\subsection{Rate models for the PDC and photoluminescence}
\label{sec:rate-model-theory}
For analysis, we model the observed rates on the detectors (``singles'') as well as the coincidences rates to simultaneously estimate the intrinsic PDC pair production rate and photoluminescence. This approach has been applied to various processes, such as PDC \cite{Pearson2010ratemodels, Schneeloch2019} or FWM \cite{Silverstone2014,Faruque2019}. In our case, the model is the system of equations,
\begin{align}
R_s &= \eta_s\left(\xi P + f(P)\right)+R_\text{bg},\label{eq:Rs}\\
R_i &= \eta_i\left(\xi P + f(P)\right)+R_\text{bg} \;\text{and}\label{eq:Ri}\\
R_{c} &= \eta_s\eta_i\xi P + \tau_cR_sR_i,\label{eq:Rsi}
%Alternative equation also includes terms (see Silke's thesis): \tau_c\left(R_s-\eta_s\eta_i\xi P\right)\left(R_i-\eta_s\eta_i \xi P\right)
\end{align}
which we solve for the detected coincidence rate $R_{c}$. Empirically, we know that the background rate $R_\text{bg}$ is equal for both channels, which we include directly in the model\footnote{In principle, as the channels are separated by polarization in the setup in Fig.~\ref{fig:main-setup}, polarized background light split at the PBS and coupled to the fibers could cause an imbalance. However, we have not observed such an effect in our system. In fact, the SNSPD bias current is set to a value that a mean dark-rate of \SI{300}{\per\second} is achieved. Thus, the quantum efficiency might be slightly different for different detectors. This is taken care of by measuring the heralding efficiencies.}. The single rates in Eqs.~\eqref{eq:Rs} and \eqref{eq:Ri} are eliminated, allowing us to write $R_{c}$ in terms of the incident optical power $P$, the efficiency to generate photon pairs from that power $\xi$, the heralding efficiencies of ``signal'' and ``idler'' $\eta_s$ and $\eta_i$ according to the Klyshko scheme \cite{klyshko1980}, and the coincidence~window~$\tau_c$. Similar to the intra-waveguide PDC generation rate $\xi P$, the noise model $f(P)$ describes the amount of photoluminescence generated depending on pump power.

A power law is the simplest noise function, which is given by
\begin{equation}
f(P)=\gamma_P P^\alpha,\label{eq:model-power-law}
\end{equation}
with $\gamma_P$ being the photoluminescence generation efficiency, equivalent to $\xi$.  This model is easy to interpret via the exponent $\alpha$ in terms of the number of photons that are involved. If it is dominated by two- or multi-photon processes or exponential avalanche effects, like in laser resonators, the exponent $\alpha$ is greater or equal than $2$. A value of $1$ corresponds to linear absorption, smaller than $1$ indicates saturation. 

The second proposed noise function describes a saturable absorber, we assume a ``dead-time'' model \cite{davidson1968} given by
\begin{equation}
f(P)=\frac{\gamma_S P}{1+\beta \gamma_S P}.\label{eq:model-dead-time}
\end{equation}
Here, $\beta$ corresponds to an effective lifetime (or dead-time) of an ensemble of light emitting defects and $\gamma_S$ corresponds to $\gamma_P$ in Eq.~\eqref{eq:model-power-law}. We focus only on these two noise models as others, like ones based on error functions $\propto \text{erf}(P)$ or exponentials $\propto 1-\exp(P)$ fail to converge satisfactorily over the whole power range for the data presented in section~\ref{sec:rate-model}. 

%----------------------------------------------------------------------------------
\section{Results}
%----------------------------------------------------------------------------------

\subsection{Spatial distribution of the photons at the facet}
\label{sec:spatial}
For start, we determine the spatial distribution of the PDC and background counts of the \SI{2.04}{\milli\meter} waveguide. This is done to verify that both coincidences and background originate at the waveguide and are not collected from somewhere else. Moreover, there have been previous hypotheses that the background results from impurities in the GaAs substrate  \cite{ Boitier-Ducci-Electrically-2013}, which can be verified by comparison with the spatial distribution of the PDC signal.

The collection fiber is held by a clamp mounted on an Elliot Scientific MDE510 fiber launch system. The fiber can be moved in X, Y, and Z-directions relative to the fixed collimating lens. We measure the distribution by moving the fiber in the imaging plane parallel to the waveguide facet. Due to the different focal lengths of the collimating lenses, the image of the waveguide facet is magnified 5.8 times at the plane of the collection fiber.

This approach is limited in resolution by the collection spot size and the difficulty to determine the absolute position of the spot on the facet. The latter can be partially circumvented, as we optimize the coupling for maximum coincidences. According to simulations, the maximum of the coincidences is the center of the ridge and the center of the core layer. We choose this easy to find position as our reference point. First, we scan the distribution of the coincidences, starting with the maximum. Then, we optimize for maximum coincidences again and change the wavelength of the pump lasers, so that no PDC can be detected. This allows us to repeat the same measurement for the photoluminescence using the same reference frame. 
% Densityplot of background:
% Fig.~\ref{fig:spatial-bg} shows the resulting distribution of the photoluminescence.

% \begin{figure}[ht]
% \centering
% \includegraphics{fig-spatial-bg.pdf}
% \caption{\label{fig:spatial-bg} Distribution of the detected photoluminescence rates in one channel at the facet. The coupling fiber is scanned parallel to the facet with the X- and Z-axis of the translation stage in the setup depicted in Fig.~\ref{fig:main-setup}.}
% \end{figure}

\begin{figure}[ht]
\centering
\includegraphics{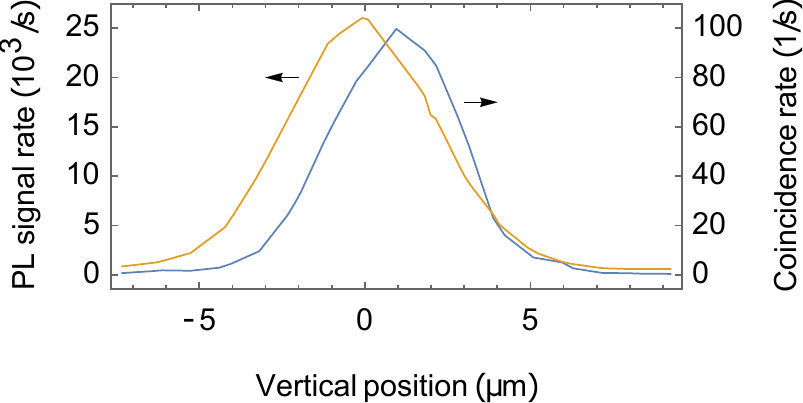}
\caption{\label{fig:spatial-pdc} Vertical slice of the photoluminescence signal (orange, left scale) and the coincidences of the PDC (blue, right scale). The peak of the background is slightly below the peak of the PDC coincidences.}
\end{figure}

In the horizontal direction, which is parallel to the epitaxial layers, both distributions are perfectly centered, but the photoluminescence is much wider.  In vertical direction, which is perpendicular to the layers of the BRW structure, the photoluminescence is also wider and the maximum is shifted by approximately \SI{1}{\micro\meter} towards the substrate, as shown in Fig.~\ref{fig:spatial-pdc}.  We observe little to no light from the substrate, even though the peak of the photoluminescence is observed slightly shifted towards the substrate. A possible explanation for the wider and shifted emission of the photoluminescence is that there are many  modes the impurities can emit into, instead of just the total internal reflection modes for PDC. They may even be only weakly guided and just diffract through the sample, but only the parts that are actually collected are relevant in further experiments.

The GaAs substrate is semi-insulating, which increases the total number of impurities and potentially the photoluminescence. In contrast to previous hypotheses about the spatial distribution \cite{Boitier-Ducci-Electrically-2013}, our data suggests that the substrate causes little or no relevant photoluminescence. This is important for electrically active samples: If a heavily doped substrate is required to contact the waveguide from below, it will not affect the noise generation rate. Nevertheless, higher doping in the core regions increase the number of impurities, so more photoluminescence has to be expected.

\subsection{Rate models from power sweeps at the degeneracy wavelength}
\label{sec:rate-model}
In the second experiment, we record the single count rates as well as the coincidences in a \SI{13}{ns} time window over three orders of magnitude of input power and compare them with the case of narrower temporal time-gating of \SI{1.13}{ns}. The power sweeps are carried out at a pump wavelength of \SI{763}{nm}, which is approximately \SI{0.3}{\nano\metre} below the degeneracy wavelength of the employed \SI{1.3}{\milli\meter} long waveguide. This wavelength was chosen empirically, as it is a stable and repeatable laser operating point. We select average pump powers with logarithmic spacing between \SI{10}{\micro\watt} and \SI{2000}{\micro\watt}, the typical operating range of our waveguides when pumped externally. 

Each power sweep corresponds to one of four conditions: (1) Pulsed pump without filter, (2) pulsed pump with \SI{40}{nm} or (3) pulsed pump with \SI{12}{nm} bandpass filter. The bandpass filters are centered at the degeneracy wavelength of the produced signal and idler photons. For condition (4), we remove the bandpass and couple a CW laser into the waveguide. The rationale behind this approach is two-fold: First, the SNSPDs are able to detect light well outside the telecom C-band, we can therefore test the effectiveness of bandpass filtering under pulsed pump conditions. We know from previous measurements that the spectrum of the photoluminescence is very broad and uniform, and that spectral filtering is effective in increasing the signal-to-noise ratio \cite{laiho--2016--uncovering}. Second, the peak power in the pulsed case is roughly $10^6$ times higher than in the CW case for the same average power. Any nonlinear effects, like two-photon absorption, should therefore be clearly discernible. 

\begin{figure}[ht]
\centering
\includegraphics{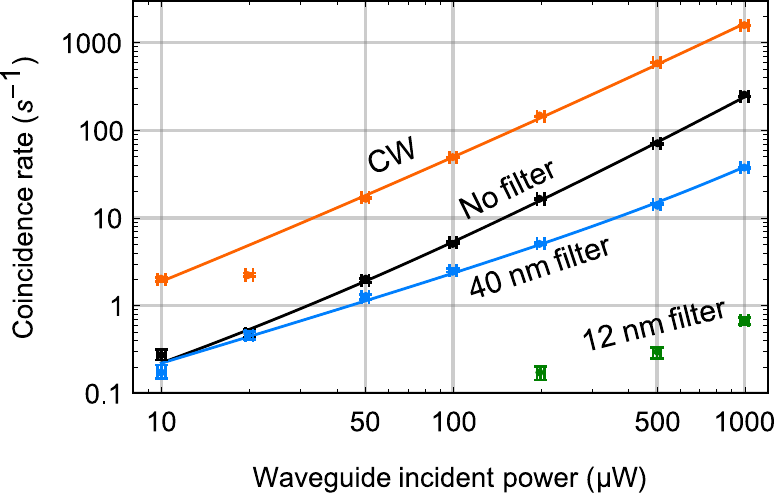}
\caption{\label{fig:rate-models} Recorded coincidences with a fit representing the power law and the saturation model for the photoluminescence using a \SI{13}{ns} window. The two models produce almost identical curves in the displayed range.}
\end{figure}

\begin{figure}[ht]
\centering
\includegraphics{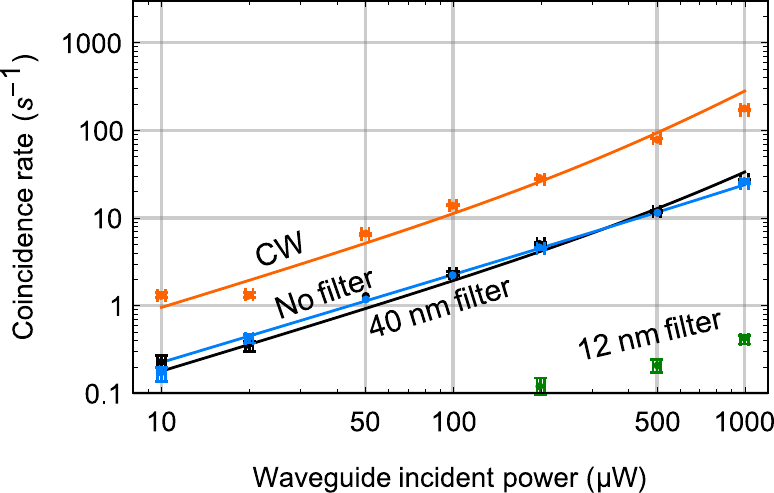}
\caption{\label{fig:time-gated-models} Recorded coincidences using a narrow \SI{1.13}{ns} window. Here, photoluminescence models are not needed ($f(P)=0$) to explain the curves statistically well enough.}
\end{figure}

The coefficients of two PDC models with different noise functions are given in table~\ref{tab:model-coefficients}, the raw data with power-law fits and with the saturation models are depicted in Fig.~\ref{fig:rate-models}. For comparison, the analysis with a narrower time gate of \SI{1.13}{ns} is shown in Fig.~\ref{fig:time-gated-models}. Here, only the rate model for the PDC without photoluminescence is fitted. Looking at the data for CW, we suspect that the data point at \SI{20}{\micro\watt} is an outlier and is subsequently excluded. Note, that the \SI{12}{nm} bandpass filter cut way too much signal to get enough data for a sound analysis.

% Second rate model pictue (but almost identical to the one above)
% \begin{figure}[ht]
% \centering
% \includegraphics{fig-coinc-saturation-models.pdf}
% \caption{\label{fig:saturation-models} Recorded coincidences with fits where the saturation model for the photoluminescence was used.  Compared to Fig.~\ref{fig:rate-models}, a slight curve at low powers, especially for the CW case, can be noted.}
% \end{figure}

\begin{table*}
\caption{\label{tab:model-coefficients} Fit coefficients of the two noise models for different pump/filter conditions according to Eqs.~\eqref{eq:Rs}-\eqref{eq:Rsi}. The data measured with the \SI{40}{nm} bandpass filter can be well explained without a noise model. Most quoted parameter values exhibit a $p$-value of lower than 0.01, except for $\xi$ in the \textit{No filter (CW)} case and $\beta$ for \textit{No filter (pulsed)} (both $p=0.1$). The $R^2$ is always better than $0.95$. The coincidence window is \SI{13.1}{ns} (= 1/laser repetition rate), except where stated.}
\begin{ruledtabular}
\begin{tabular}{llccc}
PL model & Parameter & No filter (pulsed) & \SI{40}{nm} BPF (pulsed) & No filter (CW)\\
\hline
\\[-0.5em]
Power law Eq.~\eqref{eq:model-power-law} & $\xi$ ($10^6$ pairs/s/\textmu W) & \SI{0.5(1)}{} & \SI{0.060(1)}{} & \SI{0.5(2)}{} \\
Noise $\propto \gamma_P P^\alpha$& $\gamma_P$ ($10^6$ photons/s/\textmu W) & \SI{2.0(4)}{} & n/a & \SI{8.5(5)}{} \\
& $\alpha$  & \SI{0.74(4)}{} & n/a & \SI{0.70(3)}{} \\
%Fit quality & $R^2$   & 0.998 & 0.998 & 0.998 \\
\\[-0.5em]
Saturation Eq.~\eqref{eq:model-dead-time} & $\xi$ ($10^6$ pairs/s/\textmu W) & \SI{0.78(6)}{} & \SI{0.060(1)}{} & \SI{1.40(6)}{} \\
Noise $\propto \gamma_S P/(1+\beta \gamma_S P)$& $\gamma_S$ ($10^6$  photons/s/\textmu W) & \SI{0.4(1)}{} & n/a & \SI{2.5(2)}{} \\
& $\beta$ (ns) & \SI{8(6)}{} & n/a & \SI{5(1)}{} \\
%Fit quality & $R^2$   & 0.998 & 0.998 & 0.993 \\
\\[-0.5em]
Short time gate (\SI{1.13}{ns}) & $\xi$ ($10^6$ pairs/s/\textmu W) & \SI{0.80(1)}{} & \SI{0.062(1)}{} & \SI{1.80(3)}{} \\
\\[-0.5em]
Klyshko efficiencies & $\eta_s\times10^{-4}$ & 1.9(1) & 4.9(2) & 2.6(1) \\
& $\eta_i\times10^{-4}$ & 1.5(1) & 8.5(3) & 3.2(1) \\
\end{tabular}
\end{ruledtabular}
\end{table*}

We then take a closer look at the estimated PDC pair generation rates $\xi$. As the setup and waveguide remain identical in between experiments, these values speak for the consistency of our measurements. Most of the rates are within the range of \SIrange{5e5}{8e5}{pairs\per\micro\watt\per\second} without filters, and an order of magnitude less with a \SI{40}{nm} bandpass filter. The only exception is the saturation model in the \textit{No filter (CW)} case, which reports pair rates more than twice as high as the other cases. This might be an artifact of the low dead-time it converged to. Because of the known repetition rate, we expect an effective dead-time slightly above \SI{10}{ns}. 

However, this raises the question of whether a reduction to a tenth of the PDC rate is agreeable when employing \SI{40}{nm} bandpass filter. As the spectrum of the PDC is more than \SI{90}{nm} FWHM wide \cite{Gunthner-2015}, a considerable reduction in power can already be expected. Including the transmission profile of the filter, we determine that only approximately \SI{32}{\percent} of the signal/idler photons are transmitted, while the rest is absorbed or reflected by the filter. As coincidence rates are affected twice, due to the involvement of two correlated photons, the total PDC transmission is expected to be about \SI{10}{\percent}, which is consistent with our results for the PDC rates. 

Moreover, as the noise is spectrally broader than the PDC \cite{laiho--2016--uncovering}, the signal-to-noise ratio increases. This is evident as the model without photoluminescence ($\gamma_S$ or $\gamma_P=0$) yields a statistically significant fit in Table~\ref{tab:model-coefficients}. % and Fig.~\ref{fig:car}.

% Detuning effect. Redundant, as the effect can be explained by the bandpass filter alone
%\begin{figure*}[ht]
%\centering
%\includegraphics{jsa-detuned.pdf}
%\caption{\label{fig:jsa} Effect of a detuned pump in BRWs with bandpass filtering. The plots are representative for BRWs, but slightly exaggerated for clarity. The reported values for PDC generation rate are normalized to the unfiltered case pumped at degeneracy (top left) and are the result of using realistic (experimental) parameters for each case.}
%\end{figure*}

% See above
%The remaining factor of two is most likely a result of a slightly blue-detuned pump laser, as compared to the optimal degeneracy wavelength. Since the joint spectrum is slightly curved due to non-zero the group velocity dispersion in the waveguide \cite{laiho--2016--uncovering}, pairs with vastly differing wavelength are excited with higher probability. The reasoning behind this is illustrated in Fig.~\ref{fig:jsa}. The detuning is necessary, because the lasers have to work stably for prolonged times. In our system in these wavelength bands, this is only possible at certain discreet wavelengths. In that sense, the quoted pair generation rate of \SI{0.5(1)e5}{pairs\per\micro\watt\per\s} should be seen as lower bound. 

Our results show that the exponents of the photoluminescence power law model are significantly below 1. This indicates that the driving factor of the photoluminescence is saturable linear absorption, and that higher-order photon processes play only marginal roles in our experimental conditions. In the following section~\ref{sec:off-resonant}, the dependence of the photoluminescence generation rates on the pump wavelength supports this hypothesis. The saturation is modeled as an excitation with a lifetime, which assumes that the photoluminescence stems from impurities that can only be re-excited after a delay when emitting a photon.

We emphasize that the photoluminescence scale factors $\gamma_S$ and $\gamma_P$ serve the same purpose, but their values cannot be compared directly without taking $f(P)$ into account. The true rate is only given by $f(P)$. For example, at high powers $f(P)$ predicts a much higher photoluminescence rate for the saturation model compared to the power model. At low powers it is vice-versa. Both models can explain the overall shape of the curves well and statistically sound. Further discussions about the differences are given in section~\ref{sec:future-devices} and Fig.~\ref{fig:car}.

Note, that the Klyshko efficiencies are two orders of magnitude lower as in our previous work, which is caused by the nature of this experiment. As we are deliberately trying to measure the photoluminescence, we refrain from tight spatial, temporal or spectral filtering. Hence, the single rates increase by a large amount, while the coincidence rates stay relatively low. In our previous work \cite{Gunthner-2015}, we report up to $\eta\sim \SI{6}{\percent}$ by employing the proper filters. 

\subsection{Off-resonant photoluminescence generation rates}
\label{sec:off-resonant}
In the final experiment, we move the excitation wavelength away from the degeneracy point, mostly towards longer wavelengths where no PDC is produced, because the phasematching condition is no longer fulfilled. On occasion, we have observed PDC processes of various types that probably involve higher-order modes in several waveguides. We avoid these wavelengths in this experiment to focus solely on the photoluminescence. We measure the magnitude and linearity of the photoluminescence at each wavelength for different excitation powers. 

In this section, we employ the setup with the microscope objective and the sample with a length of \SI{2.04}{\milli\meter} (degeneracy wavelength \SI{767}{nm}). We use a \SI{40}{nm} bandpass filter and a pulsed pump to emulate the conditions for a typical broadband PDC experiment with BRWs. 

\begin{figure}[ht]
\centering
\includegraphics{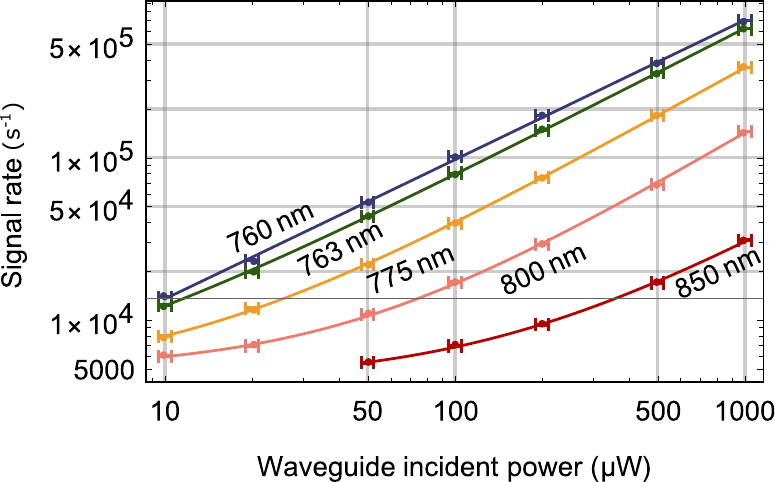}
\caption{\label{fig:off-resonant-rates} Raw data of the off-resonant photoluminescence generation rates at telecom wavelengths for different average excitation powers and pump wavelengths between \SI{760}{nm} and \SI{850}{nm}. The data is fitted with the power law from Eq.~\eqref{eq:power-law}. The fit parameters are listed in Table~\ref{tab:power-law-noise}, the generation rate $A$ is plotted in Fig.~\ref{fig:lorentzian}.}
\end{figure}

Without the PDC, we neglect coincidences so the full rate model is no longer necessary. Instead, we simply record the rates on one of the SNSPDs, measuring only dark counts plus the photoluminescence. This can be described by a power law with an offset
\begin{equation}
R_s(P)=A P^\alpha + R_\text{0},\label{eq:power-law}
\end{equation}
where $P$ is the power before the waveguide in-coupling objective. The scale factor of the polynomial $A$ can be interpreted as the photoluminescence generation rate, $R_0$ is the count rate at the lowest power measured, including dark-counts and background. The exponent $\alpha$ corresponds to the one from the fits and models discussed in the previous section~\ref{sec:rate-model}, but is determined independently for each photoluminescence data set. This model shows an excellent agreement with the measured data in Fig.~\ref{fig:off-resonant-rates}. The resulting fit coefficients are listed in table~\ref{tab:power-law-noise}. We find that $\alpha$ is very close to one, with the tendency that shorter wavelengths slightly deviate towards lower exponents. This is a further hint at saturation effects, which are more noticeable the closer the excitation wavelength is to the bandgap of the materials. We note that the value of the exponent can only be compared qualitatively between samples. The exact nature, realisation and excitation of the impurities can vary between the waveguides, resulting in slightly different rates and exponents.

\begin{table}
\caption{\label{tab:power-law-noise} Fit coefficients of the power law with offset from Eq.~\eqref{eq:power-law}.}
\begin{ruledtabular}
\begin{tabular}{@{}cccc@{}}
Wavelength & $A$ (\si{photons\per\micro\watt\per\second}) & $\alpha$ & $R_0$ (\si{photons\per\second}) \\
\hline
760\,\si{nm} & 1800(300) & 0.86(3) & \SI{100(1400)}{} \\
763\,\si{nm} & 1130(150) & 0.91(3) & \SI{3000(1000)}{} \\
775\,\si{nm} & 360(50) & 1.00(2) & \SI{4400(400)}{} \\
800\,\si{nm} & 87(13) & 1.07(3) & \SI{5000(140)}{} \\
850\,\si{nm} & 27(10) & 1.00(6) & \SI{4200(300)}{}
\end{tabular}
\end{ruledtabular}
\end{table}

\begin{figure}[ht]
\centering
\includegraphics{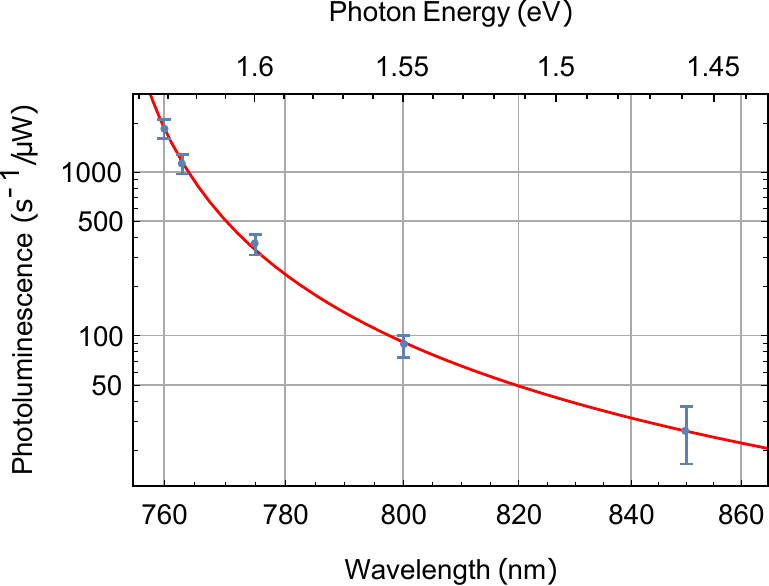}
\caption{\label{fig:lorentzian} Noise generation rate $A$ from table~\ref{tab:power-law-noise} with Lorentzian fit according to Eq.~\eqref{eq:lorentzian}. The resonance is centered at \SI{1.654(4)}{eV} (\SI{750}{nm}), which is approximately the bandgap of the matching layer materials.}
\end{figure}

The photoluminescence generation rate $A$  (table~\ref{tab:power-law-noise}) shows a distinct behaviour in Fig.~\ref{fig:lorentzian}. Over the span of \SI{100}{nm} above the bandgap, it decreases by two orders of magnitude. To model this behaviour, we tried a variety of functions, like polynomials or exponentials similar to the overlap integrals known from solid state physics \cite{gross2014}. It turns out, the only viable model is a Lorentzian function given by
\begin{equation}
A(h\nu)=\frac{\mathcal{N}}{1+\left(\frac{h\nu-E_g}{\sigma}\right)^2}.\label{eq:lorentzian}
\end{equation}
While $\mathcal{N}$ and $\sigma$ are just scale parameters, the position of the resonance, $E_g=\SI{1.654(4)}{eV} \approx \SI{750}{nm}$, can be explained physically: Its value is very close to the bandgap of the matching layers with an aluminum concentration of nominally $\SI{20}{\percent}$. The Lorentzian function is an excellent approximation for the real and imaginary parts of the dielectric function near the bandgap of Al$_x$Ga$_{1-x}$As alloys with an aluminum concentration of $x$ \cite{Kim1993dielectricalgaas}. In the literature, the quoted bandgap of Al$_x$Ga$_{1-x}$As varies slightly \cite{Kim1993dielectricalgaas,Gehrsitz-refractive-index-2000,adachi1993properties}. If we also take the typical fabrication error of the concentration of about two percentage points into account, our estimation of $E_g$ corresponds to an aluminum concentration of \SIrange{18.5}{19.5}{\percent}, which is well within the manufacturing specification. Together with the rate modeling from the previous section~\ref{sec:rate-model}, this strongly indicates that linear absorption is indeed the driving factor of the photoluminescence.

\subsection{Considerations for future devices}
\label{sec:future-devices}

Identifying and modelling the driving factors of the photoluminescence allows us to improve future samples. The first and foremost measure is increasing the spread between the lowest bandgap and the design wavelength. We propose two measures to reduce the noise: longer operating wavelengths and lower bandgap materials.
 
Quantitatively, we can estimate the effect of these proposals for the sample at hand from the fit in Fig.~\ref{fig:lorentzian}. First, moving from a \SI{767}{nm} to a \SI{780}{nm} pump wavelength already reduces the amount of photoluminescence by \SI{70}{\percent}. Second, our powerful automated sample design suite \cite{Pressl-2017-advanced-BRW} allows us to easily modify the sample to increase the aluminum concentration. For example, changing just the two matching layers next to the core from a concentration of \SI{20}{\percent} to \SI{22}{\percent}, reduces the photoluminescence by another \SI{60}{\percent}. Both measures increase the spread between excitation and bandgap edge and result in a \SI{90}{\percent} total reduction compared to the samples investigated here. It is important to note that these small changes do not affect the critical performance metrics \cite{Pressl-2017-advanced-BRW} of the waveguide, like the mode overlap and the effective nonlinearity.

Furthermore, since we now have the full rate model description (Eqs.~\eqref{eq:Rs}-\eqref{eq:Rsi} and Table~\ref{tab:model-coefficients}) of our \SI{1.3}{mm} long waveguide at hand, we calculate the  coincidence-to-accidentals ratio (CAR) curves for different filters and potential alternative designs, which may have modified material compositions, other geometries or shifted design wavelengths. A great advantage of this model-based approach is the separation of the PDC and noise signal. This allows us to evaluate the effects of different coincidence windows on the figures of merit.
 
\begin{figure*}[ht]
\centering
\includegraphics{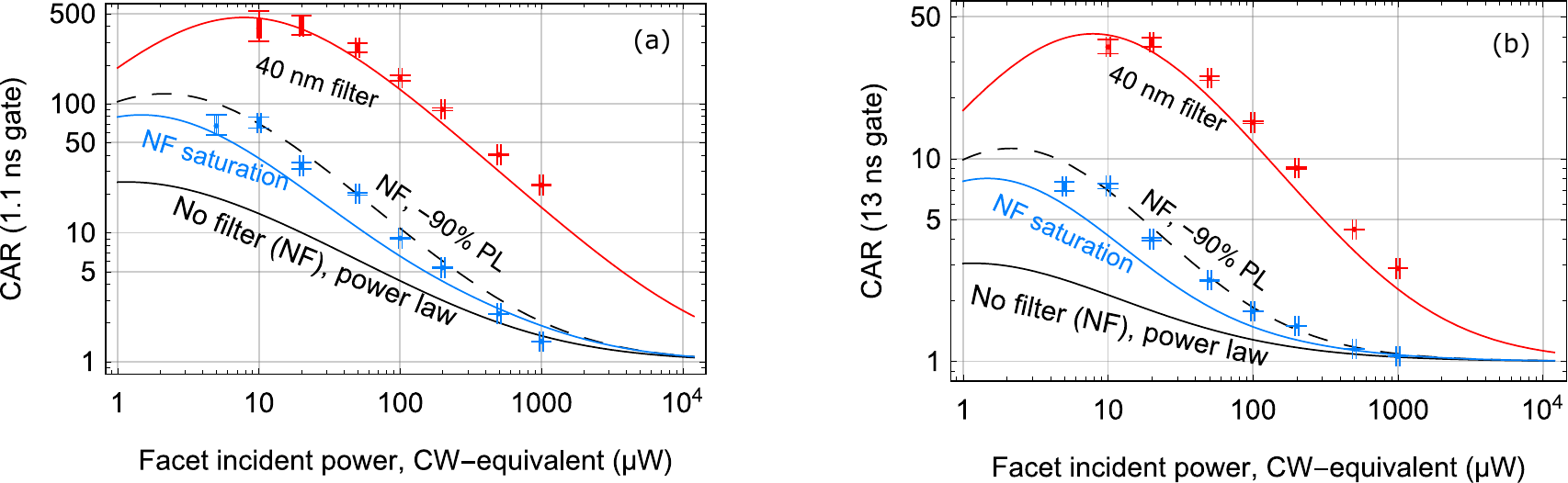}
\caption{\label{fig:car} Predicted CAR values for different conditions with a pulsed pump at \SI{76.2}{\mega\hertz} repetition rate. At low powers, the CAR is limited by the dark counts of the detectors (\SI{300}{\per\second}), at high powers by the accidental coincidences. Both the raw data (symbols) from Figs.~\ref{fig:rate-models} and \ref{fig:time-gated-models} and curves derived from the different models (\SI{40}{nm}, red and No filter, black and blue) are shown. The black solid line depicts the no filter power law model, while the blue line is the corresponding saturation model, showing better agreement to the measurements, especially at low powers. The dashed line is a hypothetical sample, with the photoluminescence reduced by \SI{90}{\percent}. Here, choosing either the power law or saturation model made no significant difference. Fig.~(a) shows the time filtered case with a \SI{1.13}{ns} time gate, (b) the unfiltered with \SI{13.1}{ns}. Note that the vertical scales are different by an order of magnitude.}
\end{figure*}

The results for two different coincidence windows are depicted in Fig.~\ref{fig:car}. It is clearly visible that spectral filtering and time filtering prove to be highly effective, as a model without photoluminescence can explain the measured data well. A narrow time gate increases the maximum CAR by a factor of 10. Without spectral filter, a noise-optimized design could increase the usable pump power substantially for a fixed CAR. The saturation model without spectral filters (solid blue line) in Fig.~\ref{fig:car} yields a CAR of 7 at \SI{100}{\micro\watt} pump power. Keeping the CAR constant and moving to the dashed line representing a sample with \SI{90}{\percent} reduction in the photoluminescence, shows a pump power of around \SI{200}{\micro\watt} - which corresponds to doubling the pair rate.

Furthermore, it is clearly visible that both noise models have their merits. We believe the power law in Eq.~\eqref{eq:model-power-law} provides a good estimate of the PDC production rate, while the saturation model in Eq.~\eqref{eq:model-dead-time} provides a good description of the processes at lower powers. The slight discrepancy with the measured data, however, also shows that we still cannot capture the full physics of our system. Moreover, the fact that two parameters in Table~\ref{tab:model-coefficients} could only be fitted to a $p$-value of 0.1 indicates the limitations of the statistics and the models in certain cases. Building the model from a pure quantum optics approach, i.e. mean photon numbers, failed to produce reliable results over the whole power range. 
 
We emphasize that the values reported in Fig.~\ref{fig:car} are conservative in the sense that we did not explicitly align for maximum CAR. Properly optimizing the pumping and coupling, also in connection with narrower filtering (e.g. \SI{12}{nm} bandpass), yields values at least an order of magnitude higher \cite{chen-2018-time-bin}. This can be seen instantly as the graph for \SI{90}{\percent} reduction is not even close to the \SI{40}{nm} case, which it should be. With better alignment, however, there is much headroom for optimizing for the individual case. Nevertheless, it follows that for these hypothetical samples, the filtering requirements are relaxed significantly compared to the state-of-the-art. This can be especially interesting for any integrated detection system: Realizing high fidelity time filtering to the picosecond level is much harder than just to nanoseconds. This is also true for on-chip spectral bandpass filters.

Having the full model depending on the input power allows us not only to estimate the CAR, but also to predict pair rates for an on-chip pump and photonic network. As all the stated input powers are measured before the aspheric lens, we need to determine the individual loss factors for coupling into the waveguide. The actual power guided in the pump mode can be recovered by multiplying the aspheric lens transmission (\SI{70}{\percent}) with the total in-coupling efficiency ($<$\SI{35}{\percent}) and the typical relative pump mode excitation ($>$\SI{4}{\percent}) \cite{Pressl--2015}. Thus, only \SI{1}{\percent} of the power reaches the necessary pump mode. Hence, we estimate that the \textit{true} coefficient of pair generation rate, e.g. for an on-chip pump, is in fact on the order of \SI{5e7}{\per\micro\watt\per\second}. In a pure externally pumped system, this is not achievable due to absorption of the glass in the objective, even with proper beam shaping to match the far field mode shape. In contrast, an active, electrically pumped, waveguide laser runs intrinsically in the correct mode \cite{bijlani2013semiconductor, Boitier-Ducci-Electrically-2013}. This means that for \SI{1}{\milli\watt} of internal laser power, a pair rate of at least \SI{5}{\giga\hertz} can be expected. Such rates are tremendously useful as they can be harnessed by a fully integrated (quantum) optic network. 
\section{Conclusion}

We have presented three different measurements designed to gain insight into the nature of BRW photoluminescence. We proposed two rate models to describe the photon generation process from a big-picture point of view. There is strong evidence that the main cause of photoluminescence is electron-hole pair excitation via linear absorption of a pump photon, followed by a short lived radiative decay via deep impurity levels. The defects that provide these deep levels are located in the matching layers with a low aluminum concentration right next to the core. Furthermore, we have proposed small modifications in the sample design that promise to greatly reduce the photoluminescence. Our calculations predict a reduction by \SI{90}{\percent}, while promising high non-linearity and photon pair rates in the GHz regime.

\section*{Author contributions}
% best practice: add author contribution statements if possible to help students write cumulative theses

%SA, BP, AS, KL and GW conceived the experiments. SA and KL built the main setup, with the assistance of AS and BP. SA, KL, AS and BP performed the measurements, SA and BP analyzed the data. HT and AS carried out refining measurements and analyses. HS, MK, SH and CS provided the samples and detailed documentation. GW supervised the project. SA and BP wrote the manuscript with the help of KL and GW. All authors contributed to the manuscript with discussions and feedback.

Conceptualization, S.A., K.L., B.P., G.W.; Formal analysis, S.A., B.P.; Methodology, S.A., A.S., K.L., B.P, G.W.; Investigation, S.A., A.S., K.L. B.P., H.T.; Resources, H.S., M.K., S.H., C.S.; Software, B.P.; Supervision, B.P., G.W.; Writing - original draft, S.A., B.P.; Writing - review \& editing, A.S., K.L., H.T., C.S., G.W.; Funding acquisition, C.S., G.W.;

\section*{Acknowledgments}
This work was supported by the Austrian Science Fund (FWF) through the project I2065 and the Special Research Program (SFB) project \textit{BeyondC} no. F7114, the DFG project no. {SCHN1376/2-1}, the ERC project {\textit{EnSeNa} (Grant No. 257531)} and EU H2020 quantum flagship program {\textit{UNIQORN} (Grant No. 820474)} and the State of Bavaria. S.A. is supported by the EU H2020 FET open project {\textit{PIEDMONS} (Grant No. 801285)}. B.P. acknowledges support by the FWF SFB project no. F6806. We thank A. Wolf and S. Kuhn for assistance during sample growth and fabrication. We thank T. G{\"{u}}nthner and H. Chen for laboratory assistance and S. Frick, M. Sassermann, R. Chapman and M. Prilm{\"{u}}ller for fruitful discussions and comments.

\section*{References}

\bibliography{brw-noise}

\end{document}